\documentclass[12pt]{article}
\usepackage[cp1251]{inputenc}
\usepackage[english]{babel}
\usepackage{amsmath,latexsym, amssymb}

\numberwithin{equation}{section}

\begin{document}

\title{Analysis of the unilateral contact problem for biphasic cartilage layers
with an elliptic contact zone and accounting for the tangential
displacements}

\author{A.A.~Koroleva\thanks{Belarusian State University, Belarus.}, \
S.V.~Rogosin\thanks{Aberystwyth University, UK, and Belarusian State University, Belarus.},
 \
G.S.~Mishuris\thanks{Aberystwyth University, UK.}
}

\maketitle

\let\thefootnote\relax\footnote{The research leading to these results has received funding from the People Programme
(Marie Curie Actions) of the European Union's Seventh Framework
Programme FP7/2007-2013/ under REA grant agreement
PIRSES-GA-2013-610547 - TAMER and by FP7-PEOPLE-2012-IAPP through
the grant PIAP-GA-2012-284544-PARM2. The authors are thankful to
Dr. I.~Argatov for important suggestions improving the proposed model.}

\noindent {\bf Abstract:} A three-dimensional unilateral contact
problem for articular cartilage layers attached to subchondral
bones shaped as elliptic paraboloids is considered in the
framework of the biphasic cartilage model.
The main novelty of the study is in accounting not only for the
normal (vertical),  but also for tangential vertical (horisontal)
displacements of the contacting surfaces.
Exact general relationships have been established between the contact approach and some integral characteristics of the contact pressure, including the contact force.
Asymptotic representations for the contact pressure integral characteristics are obtained in terms of the contact approach and some integral characteristics of the contact zone.
The main result is represented by the first-order approximation problem.

\section{Introduction}
\label{Introduction}

Biomechanical contact problems involving transmission of forces
across biological joints are of considerable practical interest
(see, e.g. \cite{Arg12,Arg13,Hunz01,Ow_Way}). Many analytical solutions to the problem of
contact interaction of articular cartilage surfaces in joints are
available. In particular, Ateshian et al. \cite{Ateshian} obtained
an asymptotic solution for the axisymmetric contact problem for
two identical biphasic cartilage layers consisting of a solid phase and a fluid phase and attached to two rigid
impermeable spherical bones of equal radii. Later, Wu et al.
\cite{Wu96} extended this solution to a more general axisymmetric
model by combining the assumption of the kinetic relationship from
classical contact mechanics \cite{John85} with the joint contact
model \cite{Ateshian} for the contact of two biphasic cartilage layers.
An improved solution for the contact of two biphasic cartilage
layers in the axisymmetric setting, which can be used for dynamic
loading, was obtained by Wu et al. \cite{Wu97}.

An asymptotic modeling approach to study the contact problem for biphasic
cartilage layers has been performed by Argatov and Mishuris in a series of articles
(see \cite{Arg_Mish_1,Arg_Mish_2,Arg_Mish_4}). In
particular,  it was shown \cite{Arg_Mish_1} that accounting for the
tangential displacements is important in the case of diseased
cartilage where the measurement of indentation depth may differ
even as much as 10\% in comparison with the healthy case. In
\cite{Arg_Mish_2}, the unilateral contact problem for articular cartilages bonded to subchondral
bones with a contact zone in the shape of an arbitrary ellipse has
been considered, and a closed form analytic solution was found.
Exploiting this exact result, Argatov and Mishuris
\cite{Arg_Mish_4} have performed perturbation analysis of the
contact problem with approximate geometry of the contact surfaces.
Other analytic solutions for the contact problem were
found using the viscoelastic cartilage model for elliptic contact
zone in \cite{Arg_Mish_3}.  A new methodology for modeling
articular tibio-femoral contact based on the developed asymptotic
model of frictionless elliptical contact interaction between thin
biphasic cartilage layers was presented in \cite{Arg12}. The
mathematical model of articular contact was extended to the case of
contact between arbitrary viscoelastic incompressible coating
layers.

In this study we extend results obtained in papers
\cite{Arg_Mish_1} and \cite{Arg_Mish_2} by considering the
influence of the tangential displacements on the contact problem
for cartilage layers with the contact zone of elliptic shape based on the
biphasic material model. Note that the perturbation method proposed in
\cite{Arg_Mish_4} could be one of the options for the analysis,
however, the procedure is too complex to perform even a few
asymptotic steps. Here, employing some technique and ideas from
\cite{Arg_Mish_1} and \cite{Arg_Mish_2}, we propose
 another way to construct the asymptotics which utilises the assumption that the
 shape of the contact zone is an ellipse at the initial stage of deformation
 and can be regarded as a small perturbation of the ellipse at any other stage of deformation.

 The paper is organized as follows. The unilateral contact problem formulation and its linearization are presented in Section~\ref{Formulation}, where a special case of the contact configuration with one cartilage layer being plane and rigid is also considered in detail.
In Section~\ref{General}, we derive exact general relationships between the contact approach and some integral characteristics of the contact pressure, including the contact force.
In Section~\ref{IAsymptoticEstimates}, we obtain asymptotic representations for the contact pressure integral characteristics in terms of the contact approach and some integral characteristics of the contact zone.
The zero-order and first-order asymptotic approximations for the solution to the contact problem are obtained in Sections~\ref{Zero-appsub} and \ref{First-appsub}, respectively. Namely, the first-order approximation problem constitutes the main result of the present study.

\section{Formulation of the contact problem}
\label{Formulation}

We consider a frictionless contact between two thin linear
biphasic cartilage layers  firmly attached to rigid bones shaped like
elliptic paraboloids. In the Cartesian co-ordinates $(x_1, x_2, z)
= ({\bf x}, z)$ the equations for the two cartilage surfaces can
be written in the form $z = (-1)^{n} \Phi^{(n)}({\bf x})$, $n = 1,
2$, where
\begin{equation}
\label{surfaces}
\Phi^{(n)}({\bf x}) = \frac{x_1^2}{2 R_1^{(n)}} + \frac{x_2^2}{2 R_2^{(n)}}
\end{equation}
with $R_1^{(n)}$, $R_2^{(n)}$ being the curvature radii of the
$n$-th bone surface at its apex.

In the undeformed  state, the cartilage-bone systems occupy convex
domains $z \leq - \Phi^{(1)}({\bf x})$ and $z \geq \Phi^{(2)}({\bf
x})$, respectively. They are in the initial contact with the plane
$z = 0$ at the origin of the co-ordinate system.

We denote by $w_1({\bf x},t)$, $w_2({\bf x},t)$ the local
vertical displacements of the corresponding cartilage surfaces. Let also ${\bf u}_1({\bf x},t)$, ${\bf u}_2({\bf
x},t)$ be the local horizontal (tangential) displacements of the
corresponding surface of the cartilages. Finally, we denote by
$P({\bf x},t)$ the contact pressure density. In this
notation the equations for the cartilage surfaces can be written
in the following form:
\begin{equation}
\begin{array}{c}
\label{pert_surfaces}
z = {\delta}_1(t) - \Phi^{(1)}\left({\bf x} + {\bf u}_1({\bf x},t)\right) + w_1({\bf x},t),\\[4mm]
z = - {\delta}_2(t) + \Phi^{(2)}\left({\bf x} + {\bf u}_2({\bf
x},t)\right) - w_2({\bf x},t).
\end{array}
\end{equation}
Here, ${\delta}_1$, ${\delta}_2$ are some (positive) vertical
displacements of the rigid bones. Note also that the vertical displacements
$w_1$, $w_2$ are positive, while the tangential displacements ${\bf
u}_1$, ${\bf u}_2$ are directed outside of the contact zone.
 Denoting by
$\delta_*(t) = {\delta}_1(t) + {\delta}_2(t)$ the contact
approach of the bones, we get
from (\ref{pert_surfaces}) the following inequality:
\begin{equation}
\label{inequality} \delta_*(t) +  w_1({\bf x},t) +  w_2({\bf x},t)
 \leq
\Phi^{(1)}\left({\bf x} + {\bf u}_1({\bf x},t)\right)
+ \Phi^{(2)}\left({\bf x} + {\bf u}_2({\bf x},t)\right).
\end{equation}

It was shown in \cite{Ateshian} (see also \cite{Arg_Mish_1}) that vertical and the tangential
displacements of each bone can be represented in the form
\begin{equation}\label{EqForW}
w_n({\bf x},t)=\frac{h_n \epsilon_n^2}{3 \mu_{s,n}} \left\{\Delta
P({\bf x},t)+\frac{3}{H_n}\int\limits_ {0}^{t}
\Delta P({\bf x},\tau) d\tau \right\},\;\quad  n = 1, 2,
\end{equation}
\begin{equation}\label{EqForU}
{\bf u}_n({\bf x},t) = - \frac{h_n \epsilon_n}{2 \mu_{s,n}}
\nabla P({\bf x},t),\;\;\; n = 1, 2.
\end{equation}
Here $\epsilon_n =h_n/a_0$ are dimensionless small parameters,
$h_1$, $h_2$ mean the thicknesses of the cartilage layers, and
$a_0$ denotes a characteristic measure of the contact zone (see
the detailed description of the role of this parameter in
\cite{Arg_Mish_1}), $H_n=(\lambda_{s,n}+2\mu_{s,n})/\mu_{s,n}$ are
material parameters of cartilages, where $\lambda_{s,n}$ and
$\mu_{s,n}$ represent the first Lame coefficient and the shear
modulus of the solid phase of the $n$-th cartilage tissue. Note
that ${\bf u}_1$ and ${\bf u}_2$ in (\ref{EqForU}) do not
necessarily coincide, they depend on both spatial variables $x_1$,
$x_2$, and on the time variable $t$.

Following \cite{Ateshian}, we introduce new spatial variables and
time variable via formulas $$x^{\prime}_j =\frac{x_j}{a_0}, \quad j =
1, 2, \quad t^{\prime} =\frac{\chi t}{\mu_0},$$ where $$\chi =
\frac{3 \mu_{s,1} k_1}{h_1^2} + \frac{3 \mu_{s,2} k_2}{h_2^2},\quad
\mu_0 = \frac{\mu_{s,1}}{\lambda_{s,1}+2\mu_{s,1}} +
\frac{\mu_{s,2}}{\lambda_{s,2}+2\mu_{s,2}},$$ $a_0$ is a
characteristic measure of the contact zone, and $k_1,  k_2$ are
the cartilage's permeabilities. In these variables we have the
following relations on the contact area $\omega(t)$ encircled by
the curve $\Gamma(t) = \partial \omega(t)$:
\begin{equation}\label{EqForW_n}
w_1({\bf x}^{\prime},t^{\prime}) + w_2({\bf x}^{\prime},t^{\prime}) =
\left(\frac{h_1^3}{3 \mu_{s,1}} + \frac{h_2^3}{3 \mu_{s,2}}\right)\left\{\Delta P({\bf x}^{\prime},t^{\prime}) + \chi \int\limits_{0}^{t^{\prime}} \Delta P({\bf x}^{\prime},\tau^{\prime}) d\tau^{\prime}\right\},
\end{equation}
\begin{equation}\label{EqForPhi}
\Phi^{(n)}({\bf x}^{\prime} + {\bf u}_n({\bf x}^{\prime},t^{\prime})) \simeq \Phi^{(n)}({\bf x}^{\prime}) - \frac{h_n^2 a_0}{2 \mu_{s,n}} \nabla \Phi^{(n)}({\bf x}^{\prime}) \cdot \nabla P({\bf x}^{\prime}, t^{\prime}), \quad n = 1, 2.
\end{equation}

Further the equality in (\ref{inequality}), i.e.,
\begin{equation}\label{EqBeforeSubst}
\delta_*(t^{\prime}) + w_1({\bf x}^{\prime},t^{\prime}) + w_2({\bf
x}^{\prime},t^{\prime}) = \Phi^{(1)}\left({\bf x} + {\bf u}_1({\bf x},t)\right)
+ \Phi^{(2)}\left({\bf x} + {\bf u}_2({\bf x},t)\right),
\end{equation}
determines the contact area $\omega(t)$.

Now we substitute (\ref{EqForW_n}), (\ref{EqForPhi}) into
(\ref{EqBeforeSubst}) and obtain the governing equation relating
the contact pressure with the vertical approach of the bones
$\delta_*(t)$ in the following form (from now on we keep the names
of new unknown functions, e.g. $\Phi({\bf x}) := \Phi({\bf
x}^{\prime} a_0)$ etc.):
\begin{equation}\label{EqForPress}
\Delta P({\bf x},t)+ {\chi} \int\limits_ {0}^{t}
\Delta P({\bf x},\tau) d\tau = {m} \left(\Phi({\bf x})-\delta_*(t)
-
\nabla \widetilde{\Phi}({\bf x}) \cdot \nabla
P({\bf x},t)\right).
\end{equation}
Here we have introduced the notation
\begin{equation}
\label{m} m = \left (\frac{h_1^3}{3 \mu_{s,1}}
+\frac{h_2^3}{3\mu_{s,2}}\right)^{-1},
\end{equation}
\begin{equation}
\label{Phi}
\Phi({\bf x}^{\prime}) = \Phi^{(1)}({\bf x}^{\prime}) + \Phi^{(2)}({\bf x}^{\prime}).
\end{equation}


Thus, it follows from (\ref{surfaces}) and  (\ref{Phi}) that the
functions $\Phi$ and $\widetilde{\Phi}$ are given by
{\begin{equation}\label{Phi_Eq} \Phi({\bf x}) = \Phi(x_1, x_2) = A
x_1^2 + B x_2^2
\end{equation}
with
$$
{A} = \frac{1}{2 R_1^{(1)}} + \frac{1}{2  R_1^{(2)}},\; \quad {B}
= \frac{1}{2 R_2^{(1)}} + \frac{1}{2 R_2^{(2)}}
$$
and
\begin{equation}\label{TildePhi_Eq}
\widetilde{\Phi}({\bf x}) =\widetilde{\Phi}(x_1, x_2) = \widetilde{A} x_1^2 + \widetilde{B} x_2^2.
\end{equation}


Note that the coefficients in $\widetilde{A}$ and $\widetilde{B} $ are positive dimensionless numbers, which are less than unit.

Without loss of generality, one can assume that $A>B$. Then, Eq. ~(\ref{EqForPress}) can
be rewritten in an equivalent form, using all dimensionless
parameters:\footnote{Note that in the axisymmetric case formula
(\ref{EqFP_new}) coinsides with formula \cite[(8)]{Arg_Mish_1}.}
\begin{equation}\label{EqFP_new}
\Delta P_\varepsilon({\bf x},t)+ {\chi} \int\limits_ {0}^{t}
\Delta P_\varepsilon({\bf x},\tau) d\tau = {\mu} \big(\Psi_1({\bf
x})-\delta_\varepsilon(t) - \varepsilon \nabla {\Psi}_2({\bf x})
\cdot \nabla P_\varepsilon({\bf x},t)\big),
\end{equation}
where the following notation has been introduced:
\begin{equation}\label{Psi_Eq}
{\Psi}_j({\bf x})= x_1^2 + e_j^2x_2^2,\quad j=1,2,\quad
\delta_\varepsilon(t)=\frac{1}{A}\delta_*(t),
\end{equation}
\[
\mu=Am,\quad e_1=\sqrt{B/A},\quad e_2=\sqrt{{\widetilde B}/{\widetilde A}},\quad \varepsilon=\frac{{\widetilde A}}{A}.
\]

It is important to note that
\begin{equation}
\label{parameters}
\chi = O(1),\quad   \mu\varepsilon\ll \chi.
\end{equation}
Discussion of the characteristic values of the introduced parameters is presented in Section \ref{Discussion} (see also \cite{Ateshian,Arg_Mish_1}).

Since the solution of (\ref{EqFP_new}) depends on the parameter
$\varepsilon$, it is customer to denote an unknown contact
pressure by $P=P_\varepsilon$ in what follows. Note that the
problem for $\varepsilon=0$ coincides with that considered in
\cite{Arg_Mish_2}, where an exact solution to this problem was
found.

{Equation (\ref{EqFP_new}) is the equation for determination of
the  contact pressure $P_\varepsilon({\bf x},t) \geq 0$, ${\bf x}
\in \omega_\varepsilon(t)$. In particular, in the case when the
contact domain is represented by an ellipse
\begin{equation}
\label{domain} \omega_\varepsilon(t) = \left\{ {\bf x}\in
{\mathbb R}^2: \; \frac{x_1^2}{b^2(t,\varepsilon)} + \frac{\beta^2(t,\varepsilon)
x_2^2}{b^2(t,\varepsilon)} \leq 1\right\}.
\end{equation}}

We supply Eq. ~(\ref{EqFP_new}) with the following boundary
conditions:
\begin{equation}\label{Bound1}
P_\varepsilon({\bf x},t)= 0,\;\;\; {\bf x}\in \Gamma(t),
\end{equation}
\begin{equation}\label{Bound2}
\frac{\partial
P_\varepsilon}{\partial n}({\bf x},t)= 0,\;\;\; {\bf x}\in \Gamma(t).
\end{equation}

The equilibrium equation
\begin{equation}
\label{load} \int\!\!\!\!\!\int\limits_{\omega_\varepsilon(t)}
P_\varepsilon({\bf x},t) d{\bf x}=F(t)
\end{equation}
connects the external load $F(t)$, unknown contact pressure
$P_\varepsilon({\bf x},t)$, and unknown contact domain  $
\omega_\varepsilon(t)$.

\subsection{Special case of the contact configuration}

In order to check the content of formula (\ref{EqForPress}) we
consider here a special case, namely, we suppose that the lower
part cartilage layer is plane and rigid (the same assumption was
employed in \cite{Wu96}), it means that $\mu_{s,2}=\infty$ and
$R_1^{(1)} = R_2^{(1)} =
\infty$, i.e.,
$$\Phi^{(1)} \equiv 0, \quad \Phi \equiv \Phi^{(2)}.
$$
In this case we have got the following equation for determination
of the contact domain  $ \omega(t)$ in the form similar to
(\ref{EqForPress}):
\begin{equation}\label{EqForPress_inf}
\Delta P({\bf x},t)+ {\chi} \int\limits_ {0}^{t}
\Delta P({\bf x},\tau) d\tau = {m} \left(\Phi({\bf x})-\delta_*(t)
-
\nabla \widetilde{\Phi}({\bf x}) \cdot \nabla
P({\bf x},t)\right).
\end{equation}
Here we will have
\begin{equation}
\label{m_inf}
m = \frac{3\mu_{s,2}}{h_2^3},\quad \chi = \frac{3 \mu_{s,2} k_2}{ h_2^2}.
\end{equation}
At the same time, small changes have to be made in the right-hand
side of Eq. ~(\ref{EqForPress_inf}) as follows:
$$
\Phi({\bf x}) = \frac{x_1^2}{2 R_1^{(2)}} + \frac{x_2^2}{2 R_2^{(2)}},
$$
$$
\tilde{\Phi}({\bf x}) = \frac{h_2^2 a_0 x_1^2}{2 \mu_{s,2} R_1^{(2)}} + \frac{h_2^2 a_0 x_1^2}{2 \mu_{s,2} R_1^{(2)}}.
$$
Thus Eq. ~(\ref{EqForPress_inf}) can be rewritten as
\begin{eqnarray}\label{EqForPress_inf1}
\Delta P({\bf x},t)  & + & \frac{3 \mu_{s,2} k_2}{ h_2^2} \int\limits_ {0}^{t}
\Delta P({\bf x},\tau) d\tau  =   \frac{3\mu_{s,2}}{h_2^3} \left(\frac{x_1^2}{2 R_1^{(2)}} + \frac{x_2^2}{2 R_2^{(2)}}
-\delta_*(t) \right) \\
{} & {} & \qquad\qquad{}-\frac{3 a_0}{h_2} \left[\frac{x_1}{R_1^{(2)}} \partial_{x_1} P({\bf x},t) + \frac{x_2}{R_2^{(2)}} \partial_{x_1} P({\bf x},t)\right].
\end{eqnarray}

It can be easily checked that in the axisymmetric case Eq.~(\ref{EqForPress_inf1}) reduces to the governing differential equation obtained in \cite{Arg_Mish_1}.

\section{General relationships between the solution components}
\label{General}
\subsection{Determination of the contact approach}

In our model we assume that the external load is non-decreasing.
Thus, the contact domain is monotonically expanded, i.e.
\begin{equation}
\label{monotone}
\omega_\varepsilon(t_1) \subseteq \omega_\varepsilon(t_2), \quad \forall t_1 \leq t_2.
\end{equation}
It is convenient to suppose also that the contact pressure is defined on the whole plane. For this we simply extend
the density $P_\varepsilon({\bf x},t)$ by assuming that
\begin{equation}
\label{pressure}
P_\varepsilon({\bf x},t) = 0,\quad \forall {\bf x}\not\in \omega_\varepsilon(t).
\end{equation}

Integrating (\ref{EqFP_new}) over contact domain
$\omega(t)$, we get
\begin{equation}\label{EqForPress1}
\begin{array}{l}
\displaystyle
\int\!\!\!\!\!\int\limits_{\omega(t)}\Delta
P_\varepsilon({\bf x},t) d {\bf x} + {\chi} \int\!\!\!\!\!\int\limits_{\omega(t)}
\!\!\int\limits_ {0}^{t} \Delta P_\varepsilon({\bf x},\tau)
d\tau d {\bf x} =
\\[5mm]
\displaystyle \hspace{20mm}
={\mu}\int\!\!\!\!\!\int\limits_{\omega(t)} \left(\Psi_1({\bf
x})-\delta_{\varepsilon}(t)\right) d {\bf x}  - {\varepsilon\mu}
\int\!\!\!\!\!\int\limits_{\omega(t)} \nabla {\Psi}_2({\bf x})
\cdot \nabla P_\varepsilon({\bf x},t) d {\bf x}.
\end{array}
\end{equation}
For simplicity of notation, we omit here (and everywhere in the
next two sections) the subindex $\varepsilon$ in
$\omega_\varepsilon$.

From the monotonicity of the contact domain (\ref{monotone}) and
assumption (\ref{pressure}), it follows that the second integral
on the left-hand side can be written in the form
\begin{equation}
\label{2+1int}
\int\!\!\!\!\!\int\limits_{\omega(t)}
\!\!\int\limits_ {0}^{t} \Delta
P_\varepsilon({\bf x},\tau) d\tau d
{\bf x} = \int\limits_{0}^{t}\!\!\int\!\!\!\!\!\int\limits_{\omega(t)} \Delta
P_\varepsilon({\bf x},\tau) d {\bf x} d\tau.
\end{equation}

Using the second Green's formula
\begin{equation}\label{Green2}
 \int\!\!\!\!\!\int\limits_{\omega(t)} \left( u({\bf x}) \Delta v({\bf x}) - v({\bf x}) \Delta u({\bf x})\right) d{\bf x} =
\int\limits_{\Gamma(t)} \left( u({\bf x}) \frac{\partial v}{\partial n}({\bf x}) - v({\bf x}) \frac{\partial u}{\partial n}({\bf x})\right)ds
\end{equation}
with $u\equiv 1$ and $v = P_\varepsilon({\bf x},t)$ we get the
following relation in view of the boundary condition
(\ref{Bound2}):
\begin{equation}
\label{int1}
\int\!\!\!\!\!\int\limits_{\omega(t)}   \Delta
P_\varepsilon({\bf x},\tau)  d {\bf x} = \int\limits_{\Gamma(t)}
\frac{\partial P_\varepsilon}{\partial n}({\bf x},s)  ds = 0,\;\;\; \forall \tau \leq t.
\end{equation}
Therefore, the both integrals on the left-hand side of
(\ref{EqForPress1}) vanish.

Further, we use the first  Green's formula
\begin{equation}
\label{Green1}
\int\!\!\!\!\!\int\limits_{\omega(t)}    \left( \varphi \Delta \psi + \nabla \varphi \cdot \nabla \psi
\right) d {\bf x} = \int\limits_{\Gamma(t)} \varphi
\frac{\partial\psi}{\partial n} ds
\end{equation}
with $\psi({\bf x}) = {\Psi}_2({\bf x})$ and $\varphi({\bf x}) =
P_\varepsilon({\bf x},t)$. In this case the integral on the
right-hand side vanishes in view of (\ref{Bound1}), and we obtain
the relation
\begin{equation}
\label{Green1-1}
\int\!\!\!\!\!\int\limits_{\omega(t)} \nabla {\Psi}_2({\bf x}) \cdot \nabla P_\varepsilon({\bf x},t) d {\bf x}
= - \int\!\!\!\!\!\int\limits_{\omega(t)} P_\varepsilon({\bf x},t) \Delta {\Psi}_2({\bf x}) d {\bf x} =
- 2 (1+e_2^2) F(t),
\end{equation}
where we used the equilibrium equation (\ref{load}) and the
identity
\begin{equation}
\label{laplace_phi_2}
\Delta {\Psi}_2({\bf x}) = 2(1 + e_2^2)
\end{equation}
with $e_2$ being defined in (\ref{Psi_Eq}).

{In what follows, it is convenient to have the following notation
for the integrals of the product of $k$-th power of the function
$\Psi_1$ and $l$-th power of the function ${\Psi}_2$:
\begin{equation}
\label{IntPhi_kl} A_{k,l}(\omega) =
\int\!\!\!\!\!\int\limits_{\omega} \Psi_1^{k}({\bf x})
{\Psi}_2^{l}({\bf x})
 d{\bf x}> 0,\;\;\; k,l = 0, 1, 2, \ldots
\end{equation}
In particular, $A_{0,0}(\omega)$ is the area of the contact
domain. It is to remember that the constants $A_{k,l}(\omega)$
depend finally on $t$, but we omitted this fact in the notation in
order to avoid cumbersome expressions. Computations of
$A_{k,l}(\omega)$ for the elliptic domain (\ref{domain}) we
included into Appendix (see Section \ref{$A_{kl}$}).}

Taking into account Eqs. (\ref{int1}) and (\ref{Green1-1}), we get
{\begin{equation}
\label{delta01} \delta_{\varepsilon}(t)=
\frac{A_{1,0}(\omega_\varepsilon(t))}{A_{0,0}(\omega_\varepsilon(t))}  + \frac{2(1 + e_2^2) \varepsilon}{
A_{0,0}(\omega_\varepsilon(t))} F(t).
\end{equation}}
This formula allows us to compute the contact approach $\delta_{\varepsilon}(t)$
as a function of the total external force $F(t)$ and the main axes
of the ellipse describing the shape of the contact zone, which in
fact depends on time too.


\subsection{Some integral identity for the contact pressure}

In order to write out a more informative equation for the contact
load, we use the following trick. We multiply   both sides of
(\ref{EqFP_new}) by the function $v({\bf x})={\Psi}_2( {\bf x})$
and integrate the obtained equation over the contact domain
$\omega(t)$
\begin{eqnarray}
 \int\!\!\!\!\!\int\limits_{\omega(t)} {\Psi}_2(
{\bf x}) \Delta P_\varepsilon({\bf x},t) d {\bf x} & + & {\chi}
\int\!\!\!\!\!\int\limits_{\omega(t)} \int\limits_ {0}^{t} {\Psi}_2( {\bf x}) \Delta
P_{\varepsilon}({\bf x},\tau) d\tau d {\bf x} = \nonumber \\
{} & = &\mu \int\!\!\!\!\!\int\limits_{\omega(t)}  {\Psi}_2( {\bf x})
\Psi_1({\bf x}) d {\bf x} - \mu {\delta_{\varepsilon}(t)}
\int\!\!\!\!\!\int\limits_{\omega(t)}  {\Psi}_2( {\bf x}) d {\bf
x} \nonumber \\
{} & {} & {}- {\mu\varepsilon}  \int\!\!\!\!\!\int\limits_{\omega(t)}
{\Psi}_2( {\bf x})  \nabla {\Psi}_2( {\bf x}) \cdot \nabla
P_\varepsilon({\bf x},t) d{\bf x}. \label{rel0}
\end{eqnarray}
Let us calculate the integrals in this relation by using Green's
formulas. For the first integral on the left-hand side we use formula
(\ref{Green2}) with $u = {\Psi}_2$, $v = P_\varepsilon$ and the
boundary conditions (\ref{Bound1}), (\ref{Bound2}). Hence, we
obtain
\begin{equation*}
\int\!\!\!\!\!\int\limits_{\omega(t)} {\Psi}_2(
{\bf x}) \Delta P_\varepsilon({\bf x},t) d {\bf x} =
\int\!\!\!\!\!\int\limits_{\omega(t)} \Delta {\Psi}_2(
{\bf x}) P_\varepsilon({\bf x},t) d {\bf x}.
\end{equation*}
Now taking into account (\ref{laplace_phi_2}), we get
\begin{equation}
\label{int2-1}
\int\!\!\!\!\!\int\limits_{\omega(t)} {\Psi}_2(
{\bf x}) \Delta P_\varepsilon({\bf x},t) d {\bf x} = 2(1+e_2^2) F(t).
\end{equation}
For the second integral on the left-hand side, we apply the same
approach, but interchange first the integrals over
${\omega_\varepsilon(t)}$ and over $\tau\in (0, t)$ exploiting the
load monotonicity. Therefore, we arrive at the equation
\begin{equation}
\label{int2-2}
\int\!\!\!\!\!\int\limits_{\omega(t)} \int\limits_ {0}^{t}  {\Psi}_2(
{\bf x}) \Delta P_\varepsilon({\bf x},\tau)  d\tau d {\bf x} =
 \int\limits_ {0}^{t} \int\!\!\!\!\!\int\limits_{\omega(t)} {\Psi}_2(
{\bf x}) \Delta P_\varepsilon({\bf x},\tau)  d\tau d {\bf x}  =
2(1+e_2^2)  \int\limits_ {0}^{t} F(\tau) d\tau.
\end{equation}
For the first and second integrals on the right-hand side, we
simply use the notation (\ref{IntPhi_kl}), which gives
\begin{equation}
\label{int2-3} \int\!\!\!\!\!\int\limits_{\omega(t)} {\Psi}_1({\bf
x}) {\Psi}_2( {\bf x})  d {\bf  x} = A_{1,1}(b; \beta),
\quad \int\!\!\!\!\!\int\limits_{\omega(t)} {\Psi}_2({\bf
x})  d {\bf x} = A_{0,1}(b; \beta).
\end{equation}
Finally, for the third  integral on the right-hand side, we make
use of the following simple formula which follows immediately from
the definition of ${\Psi}_2$:
\begin{equation*}
{\Psi}_2 \nabla {\Psi}_2 = \frac{1}{2} \nabla {\Psi}_2^2.
\end{equation*}
Then we can apply Green's formula (\ref{Green1}) and the boundary
conditions (\ref{Bound1}), (\ref{Bound2}) to find
\begin{equation*}
\int\!\!\!\!\!\int\limits_{\omega(t)} {\Psi}_2(
{\bf x})  \nabla
{\Psi}_2({\bf x}) \cdot \nabla P_\varepsilon({\bf x},t)
d{\bf x}  = - \frac{1}{2} \int\!\!\!\!\!\int\limits_{\omega(t)}
\Delta {\Psi}_2^{2}({\bf x}) P_\varepsilon({\bf x},t)
d{\bf x}.
\end{equation*}

\noindent By applying the second Green's formula (\ref{Green2})
with $u = P_\varepsilon$, $v = {\Psi}_2^2$, and the boundary
conditions (\ref{Bound1}), (\ref{Bound2}), we represent this
integral in the form
\begin{equation}
\label{int2-5}
\int\!\!\!\!\!\int\limits_{\omega(t)} {\Psi}_2(
{\bf x})  \nabla
{\Psi}_2({\bf x}) \cdot \nabla P_\varepsilon({\bf x},t)
d{\bf x} =  - \frac{1}{2} \int\!\!\!\!\!\int\limits_{\omega(t)}
{\Psi}_2^{2}({\bf x}) \Delta P_\varepsilon({\bf x},t)
d{\bf x}.
\end{equation}
This integral still contains the unknown density of contact
pressure $P_\varepsilon({\bf x},t)$. Let us define
\begin{equation}
\label{defin_M}
\mathcal{M}^{(j)}P_\varepsilon(t)\equiv
\int\!\!\!\!\!\int\limits_{\omega(t)}
{\Psi}_2^{j}({\bf x}) \Delta P_\varepsilon({\bf x},t)
d{\bf x}.
\end{equation}

 Now we rewrite the relation (\ref{rel0}) by using the results for all integrals
 (\ref{int2-1})--(\ref{int2-5}) in the following form:
{\begin{equation}
\label{rel1} 2(1+e_2^2){\mathcal K} F(t)  = \mu A_{1,1}(\omega_\varepsilon(t))
- \mu \delta_{\varepsilon}(t) A_{0,1}(\omega_\varepsilon(t)) +
\frac{\mu\varepsilon}{2} \mathcal{M}^{(2)}P_\varepsilon(t).
\end{equation}}

Here, we have introduced the Volterra operator ${\mathcal K}$ as
follows:
\begin{equation}
\label{Operator}
{\mathcal K} F(t) = F(t) + \chi \int\limits_{0}^{t} F(\tau) d\tau.
\end{equation}

Note that the integral in the right-hand side of the equation
(\ref{rel1}) allows to continue the same procedure to deliver an
asymptotic estimate for this equation.

We continue to proceed with Eq. ~(\ref{rel1}) on the next steps.

\subsection{Asymptotic estimates of the integral characteristics $\mathcal{M}^{(j)}P_\varepsilon(t)$}
\label{IAsymptoticEstimates}

Now we proceed to calculate the last integral in (\ref{rel1}). For this we multiply the governing integral equation (\ref{EqFP_new}) by ${\Psi}_2^{j}({\bf x})$ ($j\ge2$) and integrate over contact domain $\omega(t)$:
\begin{eqnarray}
 \int\!\!\!\!\!\int\limits_{\omega(t)} {\Psi}_2^{j}(
{\bf x}) \Delta P_\varepsilon({\bf x},t) d {\bf x} & + & {\chi}
\int\!\!\!\!\!\int\limits_{\omega(t)} \int\limits_ {0}^{t} {\Psi}_2^{j}( {\bf x}) \Delta
P_{\varepsilon}({\bf x},\tau) d\tau d {\bf x} = \nonumber \\
& = & \mu \int\!\!\!\!\!\int\limits_{\omega(t)}  {\Psi}_2^{j}(\overline
{\bf x}) \Psi_1({\bf x}) d {\bf x} - \mu {\delta_{\varepsilon}(t)}
\int\!\!\!\!\!\int\limits_{\omega(t)}  {\Psi}_2^{j}( {\bf x}) d
{\bf x} \nonumber \\
{} & {} & {}- {\mu\varepsilon}  \int\!\!\!\!\!\int\limits_{\omega(t)}
{\Psi}_2^{j}( {\bf x})  \nabla {\Psi}_2( {\bf x}) \cdot \nabla
P_\varepsilon({\bf x},t) d{\bf x}. \label{IntEq_j}
\end{eqnarray}
By using the same argument as on the previous step, we get
\begin{equation}
{\mathcal K} \mathcal{M}^{(j)}P_\varepsilon(t)=
\mu A_{1,j} - \mu \delta_{\varepsilon}(t) A_{0,j}(a; \beta) -
\mu\varepsilon \int\!\!\!\!\!\int\limits_{\omega(t)} {\Psi}_2^{j}(
{\bf x}) \nabla {\Psi}_2( {\bf x}) \cdot \nabla P_\varepsilon({\bf
x},t) d{\bf x}. \label{IntEq_j_c}
\end{equation}
For the last integral we use the relations
$$
{\Psi}_2^{j}(
{\bf x})  \nabla
{\Psi}_2(
{\bf x}) = \frac{1}{j+1} \nabla
{\Psi}_2^{j+1}(
{\bf x})
$$
and
$$
\int\!\!\!\!\!\int\limits_{\omega(t)} \nabla {\Psi}_2^{j+1}({\bf
x}) \cdot \nabla P_\varepsilon({\bf x},t) d{\bf x} = -
\int\!\!\!\!\!\int\limits_{\omega(t)} \Delta {\Psi}_2^{j+1}({\bf
x}) P_\varepsilon({\bf x},t) d{\bf x}.
$$
Therefore, the integral
\begin{equation}
\label{Psi_j} \mathcal{M}^{(j)}P_\varepsilon(t)=\mu
{\mathcal K}^{-1} \left\{ A_{1,j}(\omega_\varepsilon(t)) -
\delta_{\varepsilon}(t) A_{0,j}(\omega_\varepsilon(t)) +
\frac{\varepsilon}{j+1} {\mathcal K} \mathcal{M}^{(j+1)}P_\varepsilon(t)\right\}
\end{equation}
has been obtained as a solution of the integral equation (\ref{IntEq_j_c}).
Here the inverse operator ${\mathcal K}^{-1}$ is defined by the formula
\begin{equation}
\label{K_inv}
\mathcal K^{-1} Y(t) = Y(t) - \chi \int\limits_{0}^{t}
Y(\tau)e^{-\chi(t-\tau)} d\tau.
\end{equation}

Performing the same computation, we obtain the following
representation for the integral in the right-hand side of
(\ref{rel1}):
{$$
\mathcal{M}^{(2)}P_\varepsilon(t)  = \sum\limits_{j=1}^{N}
\frac{2 \varepsilon^{j-1}}{(j+1)!} \mu^{j} {\mathcal
K}^{-j} \left\{ A_{1,j+1}(\omega_\varepsilon(t)) -
\delta_{\varepsilon}(t) A_{0,j+1}(\omega_\varepsilon(t))\right\}
$$}
\begin{equation}
\label{Psi_2N}
{}+ \frac{2 \varepsilon^{N}}{(N+2)!} \mu^{N}
{\mathcal K}^{-N} \mathcal{M}^{(N+2)}P_\varepsilon(t).
\end{equation}
Substituting this representation into Eq. ~(\ref{rel1}), we finally
get
{$$
2(1+e_2^2){\mathcal K} F(t)  =
 \sum\limits_{j=0}^{N} \frac{\varepsilon^{j}}{(j+1)!} \mu^{j+1}
{\mathcal K}^{-j} \left\{ A_{1,j+1}(\omega_\varepsilon(t)) -
\delta_{\varepsilon}(t) A_{0,j+1}(\omega_\varepsilon(t))\right\}
$$}
\begin{equation}
\label{rel2}
{}+ \frac{\varepsilon^{N+1}}{(N+2)!} \mu^{N+1}
{\mathcal K}^{-N} \mathcal{M}^{(N+2)}P_\varepsilon(t),
\end{equation}
or equivalently
{$$
2(1+e_2^2){\mathcal K}^{N+1} F(t)  =
 \sum\limits_{j=0}^{N} \frac{\varepsilon^{j}}{(j+1)!} \mu^{j+1}
{\mathcal K}^{N-j} \left\{ A_{1,j+1}(\omega_\varepsilon(t)) -
\delta_{\varepsilon}(t) A_{0,j+1}(\omega_\varepsilon(t))\right\}
$$}
\begin{equation}
\label{rel2a}
{}+ \frac{\varepsilon^{N+1}}{(N+2)!} \mu^{N+1}
\mathcal{M}^{(N+2)}P_\varepsilon(t).
\end{equation}
The latter relation allows us to determine the problem parameters
asymptotically with any prescribed accuracy.

Note that apart from the fact that the shapes of the contacting
bones are elliptical paraboloids, no additional assumptions on the
shape of the contact zone have been made. On the other hand, no
proof was offered to show that the contact zone is approximately represented by
an ellipse. This will be done later.

{\bf Remark 1.} For every $t$ for which the contact pressure
$P_\varepsilon(t)$ is bounded and the contact region $\omega(t)$
belongs to a bounded domain, the remainder
$\frac{\varepsilon^{N+1}}{(N+2)!} \mu^{N+1}
\mathcal{M}^{(N+2)}P_\varepsilon(t)$ in formula (\ref{rel2a})
tends to zero as $N \rightarrow \infty$. Thus, the series
corresponding to the sum on the right hand-side of (\ref{rel2a}) is
converging.

\vspace{3mm}

\section{Asymptotic solution to the contact problem}
\label{Zero-app}

\subsection{Zero-order approximation}
\label{Zero-appsub}

First, we get solution of the problem for $\varepsilon = 0$. In
this case Eq. ~(\ref{EqFP_new}) has the form
\begin{equation}
\label{zero} \Delta P^{(0)}({\bf x}, t) + \chi \int\limits_{0}^{t}
\Delta P^{(0)}({\bf x}, \tau)  d \tau = \mu\left(\Psi_{1}({\bf x})
- \delta^{(0)}(t)\right),
\end{equation}
where $\Psi_{1}({\bf x})$ is defined in (\ref{Psi_Eq}). 
{Since we know from \cite{Arg_Mish_2} that the contact zone is an
ellipse at this stage of approximation we will have
\begin{equation}
\delta_\varepsilon = \delta^{(0)}(t) = \delta_\varepsilon(b_0(t);
\beta_0(t)) = \frac{A_{1,0}(\omega_0(t))}{A_{0,0}(\omega_0(t))}.
\label{delta_0(0)}
\end{equation}
}

Using formula (\ref{delta_0(0)}) and calculations presented in
Section \ref{$A_{kl}$} (see formula (\ref{A_{kl}_int_fin})), one
can find that
{\begin{equation}
\label{A_00} A_{0,0}(\omega_0(t)) = \frac{\pi b_0^2}{\beta_0},\quad
A_{1,0}(\omega_0(t)) =
 \frac{\pi b_0^4}{4 \beta_0^3} \left(\beta_0^2 + e_1^2\right),
\end{equation}}
and therefore
\begin{equation}
\label{delta_0_sol} \delta^ {(0)}(t) =
\frac{b_0^2\left(\beta_0^2 + e_1^2\right)}{4 \beta_0^2}.
\end{equation}
Note that formulas (\ref{A_00}) and (\ref{delta_0_sol}) contain
two known constants
 $e_1$ and $e_2$ defined in (\ref{Psi_Eq}) and two still unknown functions
 $ b_0(t)$ and $\beta_0(t)$, which are the main semi-axis and the eccentricity of the ellipse
\begin{equation}
\label{contact_0}
\omega_0(t) = \left\{ {\bf x} \in {\mathbb R}^2: \; \frac{x_1^2}{b^2_0(t)} +
\frac{\beta^2_0(t) x_2^2}{b^2_0(t)} \leq 1\right\}.
\end{equation}

The leading terms in (\ref{rel2a}) imply (for $N=0$) the following
equation:
{\begin{equation} \label{a_0(0)} 2 (1 + e_2^2) {\mathcal K} F(t) =
\mu A_{1,1}(\omega_0(t)) - \mu \delta^{(0)}(t)
A_{0,1}(\omega_0(t)).
\end{equation}}
Here, ${\mathcal K}$ is the Volterra integral operator defined in
(\ref{Operator}).

Analogously, using some results from Section \ref{$A_{kl}$}
(see, in particular, formula (\ref{A_{kl}_int_fin})), we obtain
{\begin{equation} \label{A_01} A_{0,1}(\omega_0(t)) = \frac{\pi
b_0^4}{4 \beta_0^3} \left(\beta_0^2 + e_2^2\right)
\end{equation}
and
\begin{equation}
\label{A_11} A_{1,1}(\omega_0(t)) = \frac{\pi b_0^6}{24
\beta_0^5}\left\{3  \beta_0^4 + (e_1^2 + e_2^2) \beta_0^2 + 3
e_1^2 e_2^2\right\},
\end{equation}}
and thus
\begin{equation}
\label{a_0(0)_sol} 2 (1 + e_2^2) {\mathcal K} F(t)  = \mu
\frac{\pi b_0^6}{48 \beta_0^5} \left\{3  \beta_0^4 - (e_1^2 +
e_2^2) \beta_0^2 + 3 e_1^2 e_2^2\right\}.
\end{equation}

To find the functions $ b_0(t)$ and $\beta_0(t)$ together with the
pressure distribution over the contact zone, $P^{(0)}({\bf x},t)$,
we follow \cite{Arg_Mish_2} and introduce a new unknown function
\begin{equation}
\label{new_unknown} p^{(0)}({\bf x},t) = P^{(0)}({\bf x},t) + \chi
\int\limits_{0}^{t}  P^{(0)}({\bf x}, \tau)  d \tau =
{\mathcal K} P^{(0)}({\bf x},t).
\end{equation}
In the case of monotone external load, this function should satisfy
the Poisson equation (following from (\ref{EqForPress}))
\begin{equation}
\label{Poisson} \Delta p^{(0)}({\bf x},t) = \mu\left(\Psi_{1}({\bf
x}) - \delta^{(0)}(t)\right),\quad {\bf x}\in \omega_0(t),
\end{equation}
with the boundary conditions (\ref{Bound1}),  (\ref{Bound2}).

It is customary to rewrite this relation in the form
\begin{equation}
\label{Poisson_0} G_0({\bf x}, t) = 0,
\end{equation}
where
\begin{eqnarray}
\label{G_0_0} G_0({\bf x}, t) & = & G_0(b_0, \beta_0, \delta_0) \\
{} & \equiv & \Delta p^{(0)}({\bf x},t) - \mu\left(\Psi_{1}({\bf x}) -
\delta^{(0)}(t)\right),\; {\bf x}\in \omega_0(t).
\end{eqnarray}

Bearing in mind that the function $\Psi_{1}({\bf x})$ is a
quadratic polynomial (compare with (\ref{Psi_Eq})), it is natural
to look for the solution of such problem in the form of a
polynomial in $x_1, x_2$ of the fourth degree, that is
\begin{equation}
\label{repr_0} p^{(0)}(b_0,\beta_0,\eta_0,{\bf x},t)=\eta_0(t)
\left(1 - \frac{x_1^2}{b_0^2} -  \frac{\beta_0^2
x_2^2}{b_0^2}\right) Q_0(x_1, x_2).
\end{equation}
Note that the term in the brackets vanishes on the boundary
$\omega_0$, and thus the condition (\ref{Bound1}) is satisfied
automatically.

In Section~\ref{$Q_0$}, it has been shown that $Q_0$ is a
polynomial of the second order having the form
\begin{equation}
\label{Q_0}  Q_0(x_1,x_2)=\left(1 - \frac{x_1^2}{b_0^2} -
\frac{\beta_0^2 x_2^2}{b_0^2}\right),
\end{equation}
so that
\begin{equation}
\label{repr_0_1}  p^{(0)}(x_1,x_2; t) = \eta_0(t) \left(1 - \frac{x_1^2}{b_0^2} -
\frac{\beta_0^2 x_2^2}{b_0^2}\right)^2.
\end{equation}
Taken into account this representation we arrive at the following
relations (see Section \ref{Computations_1}):
\begin{equation}
\label{ee_1_0}
\eta_0(t)=\frac{\mu\delta^{(0)}(t)}{4(1+\beta_0^2)}b_0^2,
\end{equation}
\begin{equation} \label{ee_2_0} \eta_0(t)=\frac{\mu
b_0^4}{2(6+2\beta_0^2)}= \frac{\mu b_0^4}{4(3+\beta_0^2)},
\end{equation}
\begin{equation}
\label{ee_3_0} \eta_0(t)=\frac{\mu b_0^4
e_1^2}{2(2\beta_0^2+6\beta_0^4)}= \frac{\mu b_0^4
e_1^2}{4\beta_0^2(1+3\beta_0^2)}.
\end{equation}

This system allows us to determine the unknown functions $b_0(t)$
and $\beta_0(t)$. Indeed, eliminating $\eta_0$ from the last two
equations, we get a bi-quadratic equation defining the value of
the parameter $\beta_0$, i.e.,
\begin{equation}
\label{beta_0} 3\beta_0^4 + (1 -  e_1^2) \beta_0^2 - 3 e_1^2 = 0.
\end{equation}
By definition, $\beta_0$ is a positive parameter, thus the unique
positive solution of (\ref{beta_0}) has the form
\begin{equation}
\label{beta_0_sol} \beta_0 = \sqrt{\frac{ ( e_1^2 - 1) + \sqrt{
e_1^ 4 + 34 e_1^2 + 1}}{6}}.
\end{equation}
Note that at the zero-approximation the parameter $\beta_0$ does
not depend on time. The other parameter, $\eta_0(t)$, can be
computed directly from (\ref{ee_2_0}) or (\ref{ee_3_0}), if one
knows the remaining constant $b_0(t)$. Moreover, taking into
account (\ref{ee_1_0}) and (\ref{delta_0_sol}), one can use an
equivalent formula
\begin{equation}
\label{eta_0_fin}
\eta_0(t)=\frac{\mu
b_0^4(\beta_0^2+e_1^2)}{16\beta_0^2(1+\beta_0^2)}.
\end{equation}

In the same way, one can offer, in addition to
(\ref{delta_0_sol}), two equivalent representations for the
indentation parameter
\begin{equation}
\label{delta_0}
\delta^{(0)}(t)=\frac{1+\beta_0^2}{3+\beta_0^2}b_0^2(t)=\frac{(1+\beta_0^2)e_1^2}{\beta_0^2(1+3\beta_0^2)}b_0^2(t).
\end{equation}


Finally, the major semi-axis $b_0$ of the ellipse $\omega_0$ is
determined as follows:
\begin{equation}\label{b_0(t)}
b_0(t)=\left[\left(F(t) + \chi \int\limits_{0}^{t} F(\tau) d\tau\right)
\left(\frac{ 96\beta_0^5(1+e_2^2)} {\mu\pi (3  \beta_0^4
-\beta_0^2 (e_1^2 + e_2^2) + 3 e_1^2 e_2^2)}\right)\right]^{1/6}.
\end{equation}
Note that the parameters $b_0$, $\eta_0$ as well as the
indentation, $\delta_0$, depend on time $t$ in contrast to the
ellipse eccentricity $\beta_0$.

Now, it remains only to find the pressure over the contact area.
Using (\ref{new_unknown}) and (\ref{repr_0_1}), we get
\begin{equation}
\label{P_0_K_inv}
P^{(0)}(b_0,\beta_0,\eta_0,x_1, x_2,t)={\mathcal K}^{-1} \left(\eta_0( t) Q_0(x_1,x_2)^2\right).
\end{equation}

If $(x_1,x_2)$ belongs to the initial contact zone, i.e. $1 - \frac{x_1^2}{b_0^2(t)} -  \frac{\beta_0^2
x_2^2}{b_0^2(t)}>0$, then
\begin{equation}
\label{P_0_fin1}
P^{(0)}(x_1, x_2,t)=\eta_0( t) \left(1 - \frac{x_1^2}{b_0^2(t)} -
\frac{\beta_0^2 x_2^2}{b_0^2(t)}\right)^2-\chi \int\limits_{0}^{t}
\eta_0( \tau) \left(1 - \frac{x_1^2}{b_0^2(\tau)} -
\frac{\beta_0^2 x_2^2}{b_0^2(\tau)}\right)^2e^{-\chi(t-\tau)}
d\tau.
\end{equation}
If $(x_1, x_2)$ lies outside  of the initial contact zone, i.e. $1 - \frac{x_1^2}{b_0^2(t)} -
\frac{\beta_0^2 x_2^2}{b_0^2(t)}<0$, then
\begin{equation}
\label{P_0_fin2}
P^{(0)}(x_1, x_2,t)=\eta_0( t) \left(1 - \frac{x_1^2}{b_0^2(t)} -
\frac{\beta_0^2 x_2^2}{b_0^2(t)}\right)^2-\chi
\int\limits_{t_*(x_1,x_2)}^{t} \eta_0( \tau) \left(1 -
\frac{x_1^2}{b_0^2(\tau)} - \frac{\beta_0^2
x_2^2}{b_0^2(\tau)}\right)^2e^{-\chi(t-\tau)} d\tau.
\end{equation}
The critical moment of time $t_{\ast}$ is determined by the formula
$$b_0^2(t_{\ast})=x_1^2+\beta_0^2x_2^2.$$
Using (\ref{b_0(t)}), we get
\begin{equation}
\label{F_t_ast}
F(t_{\ast})+\chi \int\limits_{0}^{t_{\ast}} F(\tau) d\tau =\frac{ \mu \pi
}{96\beta_0^5} \left(\frac{3  \beta_0^4 -\beta_0^2 (e_1^2 +
e_2^2) + 3 e_1^2 e_2^2}{1+e_2^2}\right)(x_1^2+\beta_0^2x_2^2)^3.
\end{equation}
If the load is stepwise, we have $F(t)=F_0$. Hence, we find that
\begin{equation}
\label{t_ast}
t_{\ast} = \frac{ \mu \pi  }{96\beta_0^5 \chi  F_0} \left[\frac{(3  \beta_0^4
-\beta_0^2 (e_1^2 + e_2^2) + 3 e_1^2
e_2^2)}{1+e_2^2}(x_1^2+\beta_0^2x_2^2)^3\right] - \frac{1}{\chi}.
\end{equation}

Note that in this case
\begin{equation}\label{b_0(t)_1}
b_0^6(t_{\ast})  = \frac{ 96\beta_0^5(1+e_2^2) (1 + \chi
t_{\ast})} {\mu\pi (3 \beta_0^4 -\beta_0^2 (e_1^2 + e_2^2) + 3
e_1^2 e_2^2)} F_0 .
\end{equation}



\vspace{3mm}
This finishes the zero iteration step. Note that the results of this Section after changing the notation coincide with those obtained in \cite{Arg_Mish_1}.

\subsection{First-order approximation problem}
\label{First-appsub}

For the next steps we consider an appropriately deformed contact
domain $\omega^{(1)}_{\varepsilon}$, defined as a perturbation of the zero-order
one $\omega_0$. Namely, we assume that it can be written in the form

\begin{equation}
\label{contact1} \omega^{(1)}_{\varepsilon} = \omega^{(1)}_{\varepsilon}(t) =
\Big\{(x_1,x_2): Q_0({\bf x},t)+\varepsilon Q_1({\bf x},t) \geq 0\Big\},
\end{equation}
where unknown polynomials are taken in the forms
\begin{equation}
\label{Q0} Q_0({\bf x},t)= Q_0({\bf x},\beta_1,b_1),
\end{equation}
\begin{equation}
\label{Q1} Q_1({\bf x},t)= a_{40}(t)x_1^4+a_{22}(t)x_1^2x_2^2+a_{04}(t)x_2^4.
\end{equation}

 Note that for $\varepsilon=0$ the solution form coincides with
(\ref{contact_0}), if one take $b_1\equiv b_0$, $\beta_1\equiv
\beta_0$.

The idea behind such choice of the asymptotic anzatz is to satisfy the boundary conditions (\ref{Bound1}) and (\ref{Bound2}) automatically. This will be archived by putting
\begin{equation}
\label{omega_1} P^{(1)}_{\varepsilon} = {\mathcal K}^{-1} \Big(\eta^{(1)}( t) \big( Q_0(x_1,x_2,\beta_1(t),b_1(t))+\varepsilon Q_1({\bf x},t)\big)^2\Big).
\end{equation}


Now, when the boundary conditions are valid, we will satisfy the governing equation (\ref{EqForPress}). Note that
\begin{equation}
\label{P_1st_app} P^{(1)}_{\varepsilon} = P_0 + \varepsilon P_1 + O(\varepsilon^2),
\end{equation}
where $p_j={\cal K}(P_j)$, $j=0,1$, and
\begin{equation}
\label{P_1_0}
p_0=\eta^{(1)}( t) \left(1 -
\frac{x_1^2}{b_1^2(t)} - \frac{\beta_1^2(t)
x_2^2}{b_1^2(t)}\right)^2,
\end{equation}
\begin{equation}
\label{P_1_1}
p_1=2\eta^{(1)}( t) \left(1 -
\frac{x_1^2}{b_1^2(t)} - \frac{\beta_1^2(t)
x_2^2}{b_1^2(t)}\right)Q_1({\bf x},t).
\end{equation}

Substituting this representation into Eq. ~(\ref{EqForPress}), we
obtain
\begin{equation}
\label{EqFP_new1} {\mathcal K} \left(\Delta(P^{( 0)} + \varepsilon P_1+O(\varepsilon^2))\right) = \mu \left(\Psi_{1} -
\delta_\varepsilon^{(1)} - \varepsilon \nabla
\Psi_{2} \cdot (\nabla P^{(0)} + \varepsilon \nabla
P^{(1)}+O(\varepsilon^2))\right),
\end{equation}
where the parameter $\delta_\varepsilon^{(1)}$ is represented in
the same form as $P^{(1)}_{\varepsilon}$, i.e.,
\begin{equation}
\label{delta_1st_app} \delta^{(1)}_{\varepsilon} = \delta_0 +
\varepsilon \delta_1 + O(\varepsilon^2) = \delta^{(1)} +
O(\varepsilon^2).
\end{equation}

We can write Eq. ~(\ref{EqFP_new1}) with the accuracy to the terms
of $O(\varepsilon^2)$ as follows:
\begin{equation}
\label{EqFP_new2} \Delta p^{( 0)} + \varepsilon \Delta p_1 = \mu
\left(\Psi_{1} - 
{ \delta^{(1)}} - \varepsilon
\nabla \Psi_{2} \cdot \nabla P^{(0)}\right).
\end{equation}

An extended variant of this equation can be written by using the
definition of all components of the equation and by comparing
coefficients at different powers of $x_1, x_2$, so that

\begin{equation}
\label{varepsilon_1_00} - \frac{4 \eta^{(1)}}{b_1^2} (1 +
\beta_1^2) = - \mu \delta^{(1)},
\end{equation}
\begin{equation}
\label{varepsilon_1_20} {4 \eta^{(1)}} \left[\frac{3 +
\beta_1^2}{b_1^4} + \varepsilon (6 a_{40} + a_{22})\right] = \mu
(1 - 8 \varepsilon \theta_{2,0}),
\end{equation}
\begin{equation}
\label{varepsilon_1_02} {4 \eta^{(1)}} \left[\frac{\beta_1^2 (1 +
3 \beta_1^2)}{b_1^4} + \varepsilon (a_{22} + 6 a_{04})\right] =
\mu (e_1^2 - 8 \varepsilon e_2^2 \theta_{2,2}),
\end{equation}
{\begin{equation}
\label{varepsilon_1_22} - \varepsilon\frac{24 \eta^{(1)}}{b_1^2}(a_{40}
\beta_1^2 + a_{22} (1 + \beta_1^2) + a_{04}) = 8\varepsilon \mu (1 + e_2^2)
\theta_{4,2},
\end{equation}
\begin{equation}
\label{varepsilon_1_40} - \varepsilon\frac{4 \eta^{(1)}}{b_1^2}(a_{40} (15 +
\beta_1^2) + a_{22}) = 8\varepsilon \mu \theta_{4,0},
\end{equation}
\begin{equation}
\label{varepsilon_1_04} -\varepsilon \frac{4 \eta^{(1)}}{b_1^2}(a_{04} (15
\beta_1^2 + 1) + a_{22} \beta_1^2) = 8\varepsilon \mu e_2^2 \theta_{4,4},
\end{equation}}
where
\begin{equation}
\label{theta}
\theta_{2k,2l}(t) = {\mathcal K}^{- 1} \left(\eta^{(1)} b_1^{-2 k} \beta_1^{2 l}\right),\;\;\; k, l = 0, 1, 2.
\end{equation}

In the system (\ref{varepsilon_1_00})--(\ref{varepsilon_1_04}) we
have 6 equations and 7 unknowns: $\eta^{(1)}(t),
\delta^{(1)}_{\varepsilon}, b_1(t), \beta_1(t)$, and  $a_{40},
a_{22}, a_{04}$ (coefficients of the polynomial $Q_1$). Therefore,
we have to add an extra equation to the above system, namely
{\begin{equation}
\label{varepsilon_1} \delta^{(1)}(t)=
\frac{A_{1,0}(\omega_\varepsilon(t))}{A_{0,0}(\omega_\varepsilon(t))}  + \frac{2(1 +
e_2^2) \varepsilon}{ A_{0,0}(\omega_\varepsilon(t))} F_1(t),
\end{equation}}
where $F_1(t)$ can be represented in the form
$$
F_1(t) = \int\!\!\!\int\limits_{\omega_{\varepsilon}^{(1)}}
P^{(1)}_{\varepsilon}({\bf x}, t) d {\bf x}.
$$
We also make use of Eq. ~(\ref{rel2a}) written for this
approximation step  with the accuracy of $O(\varepsilon^{2})$ in
the form
{
\begin{equation}
\label{rel2aaa}
2(1+e_2^2){\mathcal K}^{2} F(t)  =
 \sum\limits_{j=0}^{1} \frac{\varepsilon^{j}}{(j+1)!} \mu^{j+1}
{\mathcal K}^{1-j} \left\{ A_{1,j+1}(\omega_\varepsilon(t)) -
\delta^{(1)}(t) A_{0,j+1}(\omega_\varepsilon(t))\right\}.
\end{equation}}

{
{\bf Remark 2.} Note that putting $\varepsilon=0$, the system
(\ref{varepsilon_1_00})--(\ref{varepsilon_1_04}),
(\ref{varepsilon_1})  transforms to the previous case evaluated in
the previous section.}

{{\bf Remark 3.} In the case when $\varepsilon>0$, the system
(\ref{varepsilon_1_00})--(\ref{varepsilon_1_04}),
(\ref{varepsilon_1}) has to be solved numerically. Note that the
parameter $\varepsilon$ in the last three equations
(\ref{varepsilon_1_22}) -- (\ref{varepsilon_1_04}) can be
canceled. We left these multipliers here to explain the limiting case
($\varepsilon=0$).}



\section{Discussion and conclusion}
\label{Discussion}

First of all, observe that at $t=0$, the contact problem for biphasic layers reduces to that for elastic incompressible layers. The contact problem in the latter case were studied in a number of papers \cite{Alexandrov2003,Barber1990,Chadwick2002,Yang1998}, however, without taking into account the tangential displacements.

To solve the resulting problem (\ref{varepsilon_1_00})--(\ref{varepsilon_1_04})
and (\ref{varepsilon_1}), we suggest the following iterative algorithm:

\begin{itemize}
\item
Taking $\varepsilon=0$, we have computed all values
$\eta,b,\beta,\delta = $ $\eta_0,b_0,\beta_0,\delta_0$ from the
zero-order approximation.
\item Having them we can compute the quantity $\theta_{2k,2l}(t)$ from (\ref{theta}),
\item Then, from the system of three equations (\ref{varepsilon_1_22})--(\ref{varepsilon_1_04}) we compute the constants
$a_{40}, a_{22}, a_{04}$ assuming the values of $\eta,b,\beta$ as above.
\item Finally from the system of four equations (\ref{varepsilon_1_00})--(\ref{varepsilon_1_02})
and (\ref{varepsilon_1}) considering the right-hand side known
(computed by the values know from the previous computations), we
found new values $\eta,b,\beta,\delta$ and compare them with the
previous computations. If the required accuracy has achieved we
stop the computation, if not we are going to the second step of
this iterative procedure.
\end{itemize}

We note that formulas (\ref{EqForW}) and  (\ref{EqForU}) for the
vertical and tangential displacements contain different powers of
parameters $\epsilon$, namely,  $\epsilon^2$ and $\epsilon$,
respectively. Note also that our analysis (with the values of
another parameters taken into account) shows, that the role of these magnitudes
(vertical and tangential displacements) is quite opposite. In the
final equation  (see (\ref{EqFP_new})) the leading terms,
corresponding to the vertical displacement, contain the zero
power of the new small parameter $\varepsilon$, but the leading
terms, corresponding to the tangential displacements, contain the
first power of $\varepsilon$.


\section{Appendix}

\subsection{Calculation of the constants $A_{kl}$}
\label{$A_{kl}$}

Here, we compute the values of the constants
\[
A_{k,l}(b; \beta) =  \int\!\!\!\!\!\int\limits_{\omega(t)}
\Psi_1^{k}({\bf x}) {\Psi}_2^{l}({\bf x})
 d{\bf x}> 0,\;\;\; k,l = 0, 1, 2, \ldots
\]
First of all, we note that an unknown contact domain $\omega(t)$
is of the same type as sections of the initial gap elliptical
paraboloid, i.e., it is an ellipse coaxial to the ellipse
$$
\omega(t) = \omega_\varepsilon(t) = \left\{{\bf x}\in {\mathbb
R}^2: \frac{x_1^2}{b^2(t; \varepsilon)} + \frac{x_2^2 \beta^2(t;
\varepsilon)}{b^2(t; \varepsilon)} \leq 1\right\}.
$$
In order to avoid long formulas, we use the short notation for
$\omega(t)$, writing all parameters without variables they depend
on, i.e.,
$$
\omega(t) = \left\{{\bf x}\in {\mathbb R}^2:  \frac{x_1^2}{b^2} +
\frac{x_2^2 \beta^2}{b^2} \leq 1\right\}.
$$
Performing the standard change of variables
$$
x_1 = b r \cos\, \theta,\quad x_2 = \frac{b}{\beta} \sin\, \theta,
$$
we represent the integral for $A_{k,l}(t)$ in the form
\begin{eqnarray}
A_{k,l}(b; \beta) & = & \int\limits_{0}^{1}\int\limits_{0}^{2\pi}
\left(b^2 r^2  \cos^2 \theta + \frac{b^2 e_1^2}{\beta^2} r^2
\sin^2 \theta\right)^k \left(b^2 r^2  \cos^2 \theta + \frac{b^2
e_2^2}{\beta^2} r^2 \sin^2 \theta\right)^l  \frac{b^2}{\beta} r dr
d \theta \nonumber \\
 {} & = & \frac{b^{2k + 2l +2}}{\beta} \int\limits_{0}^{1}
r^{2k + 2l + 1} dr \int\limits_{0}^{2\pi} \sum\limits_{i=0}^{k}
\frac{k!}{i! (k-i)!} \frac{e_1^{2 i}}{\beta^{2 i}}\sin^{2 i}
\theta \cos^{2k - 2 i} \theta \label{A_{kl}_int} \\
{} & {} & \times \sum\limits_{j=0}^{l}
\frac{l!}{j! (l-j)!} \frac{e_2^{2 j}}{\beta^{2 j}}\sin^{2 j}
\theta \cos^{2l - 2 j} \theta  d \theta \nonumber \\
{} & = &  \frac{b^{2k + 2l +2}}{(2k + 2l +2) \beta} \int\limits_{0}^{2\pi}
\sum\limits_{i=0}^{k} \frac{k!}{i! (k-i)!} \frac{e_1^{2
i}}{\beta^{2 i}}\sin^{2 i} \theta \cos^{2k - 2 i} \theta \nonumber \\
{} & {} & \times
\sum\limits_{j=0}^{l} \frac{l!}{j! (l-j)!} \frac{e_2^{2
j}}{\beta^{2 j}}\sin^{2 j} \theta \cos^{2l - 2 j} \theta  d
\theta.
\end{eqnarray}
Since the trigonometric functions are presented here only in even
powers, then the last integration can be performed over the
interval $[0,\pi/2]$ as follows:
\begin{eqnarray}
A_{k,l}(b; \beta) & = & \frac{4 b^{2k + 2l +2}}{(2k + 2l +2) \beta}
\int\limits_{0}^{\pi/2} \sum\limits_{i=0}^{k} \frac{k!}{i! (k-i)!}
\frac{e_1^{2 i}}{\beta^{2 i}}\sin^{2 i} \theta \cos^{2k - 2 i}
\theta \nonumber \\
{} & {} & \times \sum\limits_{j=0}^{l} \frac{l!}{j! (l-j)!}
\frac{e_2^{2 j}}{\beta^{2 j}}\sin^{2 j} \theta \cos^{2l - 2 j}
\theta  d \theta \nonumber \\
\label{A_{kl}_int1}
{} & = & \frac{4 b^{2k + 2l +2}}{(2k + 2l +2) \beta} \sum\limits_{i=0}^{k} \frac{k!}{i! (k-i)!}
\frac{e_1^{2 i}}{\beta^{2 i}} \nonumber \\
{} & {} & \times\sum\limits_{j=0}^{l} \frac{l!}{j! (l-j)!}
\frac{e_2^{2 j}}{\beta^{2 j}}
\int\limits_{0}^{\pi/2} \sin^{2 i + 2 j} \theta \cos^{2k - 2 i + 2l - 2 j}
\theta  d \theta.
\end{eqnarray}
The integrals in (\ref{A_{kl}_int1}) are calculated by using formulas
\begin{equation}
\label{Gamma}
\int\limits_{0}^{\pi/2} \sin^{2 p} \theta \cos^{2 q}
\theta  d \theta = \frac{1}{2} \frac{\Gamma(p + 1/2)\Gamma(q + 1/2)}{\Gamma(p + q + 1)},\;\;\; p, q > 0,
\end{equation}
and Legendre's duplication formula for the Gamma-function
\begin{equation}
\label{duplication}
\Gamma(n + 1/2) = \frac{\sqrt{2 \pi} \Gamma(2 n)}{2^{2 n - 1/2}\Gamma(n)},\;\;\; n\in {\mathbb N},
\end{equation}
as well as the relation $\Gamma(n + 1) = n !$. Finally, we arrive
at the following representation of $A_{k,l} = A_{k,l}(b; \beta)$
valid for all ${k, l}\in {\mathbb N}_0 = {\mathbb N}\cup \{0\}$:
\begin{eqnarray}
\label{A_{kl}_int_fin}
A_{k,l} & = & \frac{2 \pi b^2 \left({b/2}\right)^{2 k + 2 l}}{\beta (2 k + 2 l + 2) (k + l)!} \sum\limits_{i=0}^{k} \frac{k!}{i! (k-i)!}
\frac{e_1^{2 i}}{\beta^{2 i}}
\\
{} & {} & \times
\sum\limits_{j=0}^{l} \frac{l!}{j! (l-j)!}
\frac{e_2^{2 j}}{\beta^{2 j}} \frac{(2 i + 2 j)! (2 k - 2 i + 2 l - 2 j)!}{(i + j)! (k -  i + l - j)!}.\nonumber
\end{eqnarray}


\subsection{Computation of the polynomial $Q_0$}
\label{$Q_0$}

In order to determine the coefficients of the polynomial
$$
Q_0(x_1, x_2) = 1 + q_{1,0} x_1 +  q_{0,1} x_2
+  q_{2,0} x_1^2 + q_{1,1} x_1 x_2  + q_{0,2} x_2^2,
$$
we need to compute the normal derivative of the unknown functions
$p^{(0)}$ (\ref{repr_0}) along the elliptic boundary $\Gamma$:
\begin{equation}
\label{partial_p_0_1}
\frac{\partial p^{(0)}}{\partial n}{\Bigl.|_{\Gamma}} = \nabla p^{(0)} \cdot \overrightarrow{n} {\Bigl.|_{\Gamma}} = \eta_0(t) \left(- \frac{2 x_1^2}{b_0^2} - \frac{2 \beta_0^4 x_2^2}{b_0^2}\right) Q_0{\Bigl.|_{\Gamma}} = 0.
\end{equation}
Here we take into account the fact that, since the contact domain
is an ellipse (\ref{contact_0}), the tangential and normal vectors
to the boundary $\Gamma=\partial \Omega$ are given by
\begin{equation}
\label{vectors}
\overrightarrow{r}= \left(-\beta_0^2 x_2,
x_1\right),\quad \overrightarrow{n}= \left( x_1,
\beta_0^2 x_2\right).
\end{equation}

Then, to satisfy the boundary condition (\ref{Bound2}) the
following equation should be valid:
\begin{equation}
\label{Q_0_Gamma}
Q_0{\Bigl.|_{\Gamma}} = 0.
\end{equation}
This, in turn, is equivalent to the representation
\begin{equation}
\label{Q_0_fin}
Q_{0}(x_1, x_2) =  \left(1  -\frac{x_1^2}{b_0^2} -\frac{\beta_0^2 x_2^2}{b_0^2}\right).
\end{equation}

\subsection{Evaluation of the ellipse parameters}
\label{Computations_1}

Since
\begin{equation}
\label{p_0_fin}
p^{(0)}({\bf x},t)= p^{(0)}(x_1, x_2, t)=  \eta_0(t) \left(1 -
\frac{x_1^2}{b_0^2} -  \frac{\beta_0^2 x_2^2}{b_0^2}\right)^2,
\end{equation}
we have
\begin{equation}
\label{partial_p_0_1f}
\frac{\partial p^{(0)}}{\partial x_1} = 2\eta_0(t) \left(1 -
\frac{x_1^2}{b_0^2} -  \frac{\beta_0^2 x_2^2}{b_0^2}\right) \cdot
\left(- \frac{2 x_1}{b_0^2}\right),
\end{equation}
$$
\frac{\partial^2 p^{(0)}}{\partial x_1^2} = 2 \eta_0 \left[-
\frac{2}{b_0^2} \left(1 - \frac{x_1^2}{b_0^2} -
\frac{\beta_0^2 x_2^2}{b_0^2}\right) + \frac{2
x_1}{b_0^2}\frac{2 x_1}{b_0^2}\right].
$$
Therefore, by straightforward computations, we find that
\begin{equation}
\label{partial_p_0_12f}
\frac{\partial^2 p^{(0)}}{\partial x_1^2} = 2\eta_0 \left[-
\frac{2}{b_0^2} +  \frac{6 x_1^2}{b_0^4} +  \frac{2 \beta_0^2
x_2^2}{b_0^4}\right].
\end{equation}
\begin{equation}
\label{partial_p_0_2f}
\frac{\partial p^{(0)}}{\partial x_2} = 2 \eta_0 \left(1 -
\frac{x_1^2}{b_0^2} -  \frac{\beta_0^2 x_2^2}{b_0^2}\right) \cdot
\left(- \frac{2 \beta_0^2 x_2}{b_0^2}\right),
\end{equation}
$$
\frac{\partial^2 p^{(0)}}{\partial x_2^2} = 2 \eta_0 \left[-
\frac{2 \beta_0^2}{b_0^2} \left(1 - \frac{x_1^2}{b_0^2} -
\frac{\beta_0^2 x_2^2}{b_0^2}\right) + \frac{2  \beta_0^2
x_2}{b_0^2}\frac{2  \beta_0^2 x_2}{b_0^2}\right].
$$
Thus, we obtain
\begin{equation}
\label{partial_p_0_22f}
\frac{\partial^2 p^{(0)}}{\partial x_2^2} = 2 \eta_0(t) \left[-
\frac{2 \beta_0^2}{b_0^2} +  \frac{2 \beta_0^2 x_1^2}{b_0^4} +
\frac{6 \beta_0^4 x_2^2}{b_0^4}\right].
\end{equation}
Substituting (\ref{partial_p_0_12f}) and (\ref{partial_p_0_22f}) into the main equation
\begin{equation}
\label{G_0}
G_0(b_0,\beta_0,\delta_0)\equiv \Delta p^{(0)}({\bf x},t) - \mu \left(\Psi_1({\bf x}) -
\delta^{(0)}(t)\right)=0,
\end{equation}
where
$$
G_0=2\eta_0(t) \left[(-2)\frac{1 + \beta_0^2}{b_0^2} +
\left(\frac{6 + 2 \beta_0^2}{b_0^4} \right) x_1^2 +  \left(\frac{6
\beta_0^4 + 2 \beta_0^2}{b_0^4} \right) x_2^2\right] - \mu
\left(\Psi_1(x_1,x_2) - \delta^{(0)}(t)\right),
$$
and taking into account that
$$
\Psi_1({\bf x}) = \Psi_1(x_1,x_2) = {x_1^2} + {e_1^2 x_2^2},
$$
one concludes that the expression for $G_0$ is represented by a
second order polynomial with respect to the independent variables
$x_1$ and $x_2$ in the following form:
\begin{equation}
\label{G_0a}
G_0(b_0,\beta_0,\eta_0,\delta^{(0)})=q_0(b_0,\beta_0,\eta_0,\delta^{(0)})
+ q_1(b_0,\beta_0,\eta_0)x_1^2+q_2(b_0,\beta_0,\eta_0)x_2^2.
\end{equation}
Here the coefficients are defined as follows:
\begin{equation}
\label{p_0_equ1}
 q_0(b_0,\beta_0,\eta_0,\delta^{(0)})=\frac{4\eta_0}{\mu b_0^2} (1 +
\beta_0^2)- \delta^{(0)},
\end{equation}
\begin{equation}
\label{p_0_equ2}
 q_1(b_0,\beta_0,\eta_0)=\frac{4\eta_0}{b_0^4}(3+\beta_0^2) - \mu ,
\end{equation}
\begin{equation}
\label{p_0_equ3}
 q_2(b_0,\beta_0,\eta_0)=\frac{4\eta_0\beta_0^2}{b_0^4}(1+3\beta_0^2) - \mu  e_1^2.
\end{equation}

\subsection{Auxiliary computation}
\label{Computations_2}

Taking into account (\ref{P_1_0}), we can represent $p_0({\bf x}, t)$
in the form
\begin{equation}
\label{p_0_1}
p_0({\bf x}, t) = \eta^{(1)}(t) \left(1 - \frac{2 x_1^2}{b_1^2} - \frac{2 \beta_1^2 x_2^2}{b_1^2} + \frac{2 \beta_1^2 x_1^2 x_2^2}{b_1^4}
+ \frac{x_1^4}{b_1^4} + \frac{\beta_1^4 x_2^4}{b_1^4}\right).
\end{equation}
Hence, applying the Laplace equation, we get
\begin{equation}
\label{p_0_2}
\Delta p_0({\bf x}, t) = \eta^{(1)}(t) \left(-\frac{4}{b_1^2}(1 + \beta_1^2) + x_1^2 \frac{4}{b_1^4}(3 + \beta_1^2)
+ x_2^2 \frac{4 \beta_1^2}{b_1^4}(1 + 3 \beta_1^2)\right).
\end{equation}

Next, by using representation (\ref{P_1_1}), we can write
$p_1({\bf x}, t)$ in the form
\begin{eqnarray}
\label{p_1_1}
p_1({\bf x}, t) & = & 2 \eta^{(1)}(t) \left(a_{40} x_1^4 + a_{22} x_1^2 x_2^2 + a_{04} x_2^4  - \frac{a_{40} x_1^6}{b_1^2} - \frac{a_{22} x_1^4 x_2^2}{b_1^2} - \frac{a_{04} x_1^2 x_2^4}{b_1^2} \right.\\
{} & {} &
\left. {}- \frac{a_{40} \beta_1^2 x_1^4 x_2^2}{b_1^2} - \frac{a_{22} \beta_1^2 x_1^2 x_2^4}{b_1^2}
- \frac{a_{04} \beta_1^2 x_2^6}{b_1^2}\right).
\nonumber
\end{eqnarray}
Therefore, we obtain
\begin{eqnarray}
\label{p_1_2}
\Delta p_1({\bf x}, t) & = & 2 \eta^{(1)}(t) \biggl((12 a_{40} + 2 a_{22}) x_1^2 + (2 a_{22} + 12 a_{04}) x_2^2
 \\
{} & {} &
 {}- \frac{12 \beta_1^2 a_{40} + 12 a_{22} (1 + \beta_1^2) + 12 a_{04}}{b_1^2} x_1^2 x_2^2 \nonumber \\
{} & {} & {}- \frac{a_{40}(30  + 2 \beta_1^2) + 2 a_{22}}{b_1^2} x_1^4 - \frac{2 a_{22} \beta_1^2  + a_{04}(2  + 30 \beta_1^2)}{b_1^2} x_2^4\biggr).
\end{eqnarray}

We also use the following representations:
$$
\Psi_j({\bf x}) = x_1^2 + e_j^2 x_2^2,\;\;\; j = 1, 2.
$$
Thus, applying the gradient operator, we simply get
$$
\nabla \Psi_2 ({\bf x}) = \left(2 x_1, 2 e_2^2 x_2\right)
$$
and
$$
\nabla P_0({\bf x}, t) = \left({\mathcal K}^{- 1} \nabla p_0({\bf x}, \cdot)\right)(t).
$$
It yields the following representation:
\begin{equation}
\label{nabla_0}
\nabla \Psi_2 ({\bf x}) \cdot \nabla P_0({\bf x}, t) = - 8 \left({\mathcal K}^{- 1} \left[\eta^{(1)} \left(1 - \frac{x_1^2}{b_1^2} - \frac{\beta_1^2 x_2^2}{b_1^2}\right) \left(\frac{x_1^2}{b_1^2} + \frac{e_2^2 \beta_1^2 x_2^2}{b_1^2}\right)\right]\right)(t)
\end{equation}
$$
= - 8 x_1^2 \left({\mathcal K}^{- 1}\left(\frac{\eta^{(1)}}{b_1^2}\right)\right)(t) - 8 e_2^2 x_2^2 \left({\mathcal K}^{- 1}\left(\frac{\eta^{(1)} \beta_1^2}{b_1^2}\right)\right)(t) + 8 x_1^4 \left({\mathcal K}^{- 1}\left(\frac{\eta^{(1)}}{b_1^4}\right)\right)(t)
$$
$$
+ 8 (1 + e_2^2) x_1^2 x_2^2 \left({\mathcal K}^{- 1}\left(\frac{\eta^{(1)} \beta_1^2}{b_1^4}\right)\right)(t) + 8 e_2^2 x_2^4 \left({\mathcal K}^{- 1}\left(\frac{\eta^{(1)} \beta_1^4}{b_1^4}\right)\right)(t)
$$
$$
=: - 8 x_1^2 \theta_{2,0}(t) - 8 e_2^2 x_2^2 \theta_{2,2}(t) + 8 x_1^4 \theta_{4,0}(t) + 8 (1 + e_2^2) x_1^2 x_2^2 \theta_{4,2}(t)
+ 8 e_2^2 x_2^4 \theta_{4,4}(t).
$$
Here we have introduced the notation
$$
\theta_{2k,2l} = \left({\mathcal K}^{- 1}\left(\eta^{(1)} b_1^{-2k} \beta_1^{2l}\right)\right)(t).
$$

Combining the above results we obtain the system of equations (\ref{varepsilon_1_00})--(\ref{varepsilon_1_04}).


\end{document}